\documentclass[%
 reprint,
superscriptaddress,
 amsmath,amssymb,
 aps,
floatfix,
]{revtex4-2}

\makeatletter
\def\pdfstartlink@attr{}
\makeatother

\usepackage{graphicx}
\usepackage{bm}
\usepackage{mathtools}
\usepackage{hyperref}
\usepackage{amsmath}
\usepackage{mathrsfs}
\usepackage{xcolor}
\usepackage{soul}
\usepackage{orcidlink}

\usepackage{hyperref}
\hypersetup{
 pdftitle={Kilonova Light-Curve Interpolation with Neural Networks}, 
 pdfauthor={Peng}, 
 colorlinks=true,
 citecolor={blue},
 urlcolor={blue}
}

\definecolor{newblue}{rgb}{0.03, 0.65, 0.85}

\newcommand\edited[1]{{\color{black}#1}}

\newcommand\E[1]{\left\langle #1 \right\rangle}
\newcommand\editprr[1]{{\color{black}#1}}

\def\RIT{Center for Computational Relativity and Gravitation, Rochester Institute of Technology, Rochester, New York
 14623, USA}
\def\XCP{Computational Physics Division, Los Alamos National Laboratory, Los Alamos, NM, 87545, USA}
\def\CTA{Center for Theoretical Astrophysics, Los Alamos National Laboratory, Los Alamos, NM 87545, USA}
\def\CCSS{Computer, Computational, and Statistical Sciences Division, Los Alamos National Laboratory, Los Alamos, NM
 87545, USA}
\def\TD{Theoretical Division, Los Alamos National Laboratory, Los Alamos, NM 87545, USA}
\def\UA{The University of Arizona, Tucson, AZ 85721, USA}
\def\NM{Department of Physics and Astronomy, The University of New Mexico, Albuquerque, NM 87131, USA}

\def\GWU{The George Washington University, Washington, DC 20052, USA}
\def\UR{Department of Physics and Astronomy,  University of Rochester, Rochester, NY 14627, USA}
\def\CFA{Center for Astrophysics \(|\) Harvard \& Smithsonian, 60 Garden Street, Cambridge, MA 02138, USA}

\graphicspath{{figures/}}

\begin{document}

\bibliographystyle{unsrtnat}

\title{Kilonova Light-Curve Interpolation with \edited{Neural Networks}}

\author{Yinglei Peng\,\orcidlink{0000-0001-9438-7864}}
\affiliation{\UR}
\author{Marko Risti\'c\,\orcidlink{0000-0001-7042-4472}}
\author{Atul Kedia\,\orcidlink{0000-0002-3023-0371}}
\author{Richard O'Shaughnessy\,\orcidlink{0000-0001-5832-8517}}
\affiliation{\RIT}
\author{Christopher~J. Fontes\,\orcidlink{0000-0003-1087-2964}}
\affiliation{\CTA}
\affiliation{\XCP}
\author{Chris~L.~Fryer\,\orcidlink{0000-0003-2624-0056}}
\affiliation{\CTA}
\affiliation{\CCSS}
\affiliation{\UA}
\affiliation{\NM}
\affiliation{\GWU}
\author{Oleg Korobkin\,\orcidlink{0000-0003-4156-5342}}
\affiliation{\CTA}
\affiliation{\TD}
\author{Matthew R. Mumpower\,\orcidlink{0000-0002-9950-9688}}
\affiliation{\CTA}
\affiliation{\TD}
\author{V. Ashley Villar\,\orcidlink{0000-0002-5814-4061}}
\affiliation{\CFA}
\author{Ryan~T. Wollaeger\,\orcidlink{0000-0003-3265-4079}}
\affiliation{\CTA}
\affiliation{\CCSS}

\date{\today}

\begin{abstract}
Kilonovae are the electromagnetic transients created by the radioactive decay of freshly synthesized elements in the environment surrounding a neutron star merger. To study the fundamental physics in these complex environments, kilonova modeling requires, in part, the use of radiative transfer simulations. The microphysics involved in these simulations results in high computational cost, \edited{prompting the use of emulators for parameter inference applications}. Utilizing a training set of 22248 high-fidelity simulations \editprr{(composed of 412 unique ejecta parameter combinations evaluated at 54 viewing angles)}, we use a neural network to efficiently train on existing radiative transfer simulations and predict light curves for new parameters in a fast and computationally efficient manner. Our \edited{neural network} can generate millions of new light curves in under a minute. We discuss our emulator's degree of off-sample reliability and parameter inference of the AT2017gfo observational data. Finally, we \edited{discuss tension introduced by multi-band inference in the parameter inference results}, particularly with regard to the \edited{neural network's} \edited{recovery} of viewing angle.
\end{abstract}

\maketitle

\section{Introduction}
\label{sec:intro}

On August 18, 2017, prompt observations identified  gravitational wave emission (GW170817; \cite{LIGO-GW170817-bns,
  LIGO-GW170817-mma}), shortly followed by a gamma-ray burst (short-GRB GRB170817A \cite{LIGO-GW170817-grb}).  Extensive followup
observations identified a long-duration optical/near-infrared counterpart, AT2017gfo, \edited{later identified as a ``kilonova"} 
\citep{2017PASA...34...69A,2017Natur.551...64A,2017Sci...358.1556C,2017Sci...358.1570D,2017ApJ...848L..17C,Diaz17,Evans_2017,2017SciBu..62.1433H, 2017Sci...358.1559K, 2017ApJ...850L...1L,2017Natur.551...67P,2017Natur.551...75S,2017ApJ...848L..27T, Troja_2017,2017PASJ...69..101U,2017ApJ...848L..24V,2018ApJ...852L..30P, shappee2017early}.
\edited{A kilonova \cite{2010MNRAS.406.2650M} is characterized by thermal emission from rapidly expanding, radioactively-powered, heavy-element material
ejected from the associated progenitor merger.} 
The detection of the joint gravitational- and electromagnetic-wave emission from GW170817 and AT2017gfo has initiated an era of precision kilonova observations. 

Most interpretations of kilonova observations, \editprr{including kilonova candidates contributing to excess optical/infrared emission in GRB afterglows \cite{2009ApJ...696.1871P, 2013Natur.500..547T, 2014ApJ...780..118F, 2015ApJ...811L..22J, 2016NatCo...712898J, 2016ApJ...833..151F, 2017ApJ...843L..34K},} have relied on broadband photometry. \editprr{This reliance was} in part owing to the relative sparsity of available spectra for AT2017gfo (and lack of spectral observations of other kilonovae; although see e.g. \cite{GillandersNature2023, YangNature2023, LevanNature2023})
\cite{2019MNRAS.486..672A, 2013Natur.500..547T, 2014ApJ...780..118F, Pandey2019, Troja2018, Lamb2019, Troja2019, 2017ApJ...843L..34K, 2018ApJ...857..128J, 2020NatAs...4...77J, 2016NatCo...712898J, 2007ApJ...662.1129O, 2015NatCo...6.7323Y, 2015ApJ...811L..22J, 2007ApJ...655L..25Z, 2022Natur.612..228T, 2022Natur.612..223R, 2022Natur.612..232Y}. A comprehensive review of kilonova broadband photometry has recently been compiled and presented in Ref. \cite{Troja2023}. Many studies interpreted their observations of AT2017gfo shortly
after detection principally by comparison  to simplified  models for kilonovae \citep{Villar_2017, 2017ApJ...848L..17C, 2017ApJ...850L..37P,
2017PASA...34...69A, 2017Natur.551...64A, 2017Sci...358.1556C, 2017Sci...358.1570D, 2017ApJ...848L..17C, Diaz17, Evans_2017, 2017Sci...358.1559K, 2017Natur.551...67P, 2017Natur.551...75S, 2017ApJ...848L..27T, Troja_2017, 2017PASJ...69..101U, 2017ApJ...848L..24V, 2018ApJ...852L..30P, shappee2017early}
consisting of one or more groups of non-accelerating (homologous) expanding material. 
Motivated both by binary merger simulations and the inability to fit observations with one component \cite{2017ApJ...848L..17C}, at least two
components are customarily employed \cite{Evans_2017, 2017Sci...358.1559K, 2017Natur.551...75S, 2017ApJ...848L..27T, Troja_2017, 2018ApJ...852L..30P}, with properties loosely associated with two expected features of merger
simulations: promptly ejected material (the ``dynamical'' ejecta), associated with tidal tails or shocked
material at contact; and material driven out on longer timescales by properties of the remnant system (the ``wind'' ejecta) \citep{2019ARNPS..69...41S}.

Radiative transfer models of two-component kilonovae, whilst more physically accurate and informative, come with a significantly higher computational cost \edited{compared to semi-analytical or one-component models}. As a result of this cost, many groups have resorted to surrogate models, or emulators, for the kilonova outflow, to reduce the computational impact associated with inference using these more complex models \citep{gwastro-mergers-em-CoughlinGPKilonova-2020, 2021arXiv211215470A, Ristic22, 2022MNRAS.516.1137L, 2023arXiv230711080A, kilonovAE,  2023PhRvR...5d3106R, 2023PhRvR...5a3168K, 2023arXiv231109471D, 2024ApJ...961..165S}. 

In this paper, we use a neural network emulator for light-curve interpolation and parameter inference. \edited{We train on a previously-generated library of $\sim 400$ two-component kilonova light-curve simulations.} Our method can be easily applied to any modestly-sized archive of adaptively-learned astrophysical transient light-curve simulations. 
\editprr{Like our previous study \cite{Ristic22}, our investigation is a concrete and novel realization of a widely-used strategy in many areas of physics: training a
  surrogate model against reference simulations, then using the surrogate for model parameter inference.   In this work,
  however, we perform a novel investigation to assess the quality of our surrogate, to find surprising
  differences relative to conclusions obtained with the previously-described surrogate \cite{Ristic22}.  Our
results  raise  general questions about how to appropriately characterize and propagate  systematic uncertainty in
surrogate models, which will be of broad interest in any area of physics where comparable surrogate approaches are used.
}

This paper is organized as follows. In Sec. \ref{sec:intp} we discuss our simulation training library, the interpolation model and its architecture, and the associated light-curve interpolation methodology. In Sec. \ref{sec:results} we compare our emulator's performance with others employed in the literature and report inference results for observations of AT2017gfo. In Sec. \ref{sec:conclusion}, we summarize our findings.

\section{Interpolation Methodology}
\label{sec:intp}

\subsection{Simulation Description}
\label{sec:sim_setup}
Our two-component kilonova model consists of a lanthanide-rich equatorial dynamical ejecta component and a lanthanide-poor axial wind ejecta component as described in \cite{kilonova-lanl-WollaegerNewGrid2020, 2021ApJ...910..116K} and motivated by numerical simulations \citep{2019ARNPS..69...41S,just23}. Each component is homologously expanding and parameterized by a mass and velocity such that $M_{\rm{d}}$, $v_{\rm{d}}$ and $M_{\rm{w}}$, $v_{\rm{w}}$ describe the dynamical and wind components' masses and averaged velocities, respectively. The morphology for the dynamical component is an equatorially-centered torus, whereas the wind component is represented by an axially-centered peanut component; Figure~1 of \cite{kilonova-lanl-WollaegerNewGrid2020} displays the torus-peanut, or ``TP," schematic corresponding to the morphologies employed in this work \citep[see][for detailed definition]{2021ApJ...910..116K}. The lanthanide-rich dynamical ejecta is a result of the $r$-process nucleosynthesis from a neutron-rich material with a low electron fraction ($Y_{\rm{e}} \equiv n_{\rm{p}}/(n_{\rm{p}} + n_{\rm{n}})$) of $Y_e = 0.04$ with elements reaching the third $r$-process peak ($A \sim 195$), while the wind ejecta originates from higher $Y_e = 0.27$ which encapsulates elements between the first ($A \sim 80$) and second ($A \sim 130$) $r$-process peaks. The detailed breakdown of the elements in each component can be found in Table~2 of Ref.~\cite{kilonova-lanl-WollaegerNewGrid2020}.

We use \texttt{SuperNu} \citep{SuperNu}, a Monte Carlo code for simulation of time-dependent radiation transport with matter in local thermodynamic equilibrium, to create simulated kilonova spectra $F_{\lambda, \rm sim}$ assuming the aforementioned two-component model. Both components are assumed to have fixed composition and morphology for the duration of each simulation. \texttt{SuperNu} uses radioactive power sources calculated from decaying the $r$-process composition from the \texttt{WinNet} nuclear reaction network \citep{2012ApJ...750L..22W,Korobkin_2012,Reichert2023a,Reichert2023b}. These radioactive heating contributions are also weighted by thermalization efficiencies introduced in Ref.~\cite{Barnes_2016} (see Ref. \cite{Wollaeger2018} for a detailed description of the adopted nuclear heating). We use detailed opacity calculations via the tabulated, binned opacities generated with the Los Alamos suite of atomic physics codes \citep{2015JPhB...48n4014F,2020MNRAS.493.4143F,nist_lanl_opacity_database}. In the database that we use, the tabulated, binned opacities are not calculated for all elements; therefore, we produce opacities for representative proxy elements by combining pure-element opacities of nuclei with similar atomic properties \citep{2020MNRAS.493.4143F}. Specifics of the representative elements for our composition are given in Ref.~\cite{kilonova-lanl-WollaegerNewGrid2020}.

The \texttt{SuperNu} outputs are observing-angle-dependent, simulated spectra $F_{\lambda, \rm sim}$, post-processed to a source distance of $10$ pc, in units of erg s$^{-1}$ cm$^{-2}$ \AA$^{-1}$. The spectra are binned into 1024 equally log-spaced wavelength bins spanning $0.1 \leq \lambda \leq 12.8$~microns.

For the purposes of this work, we consider the light curves for the 2MASS $grizy$ and \edited{Rubin Observatory} $JHK$ broadband filters. As we only consider anisotropic simulations in this study, unless otherwise noted, we extract simulated light curves using 54 angular bins, uniformly spaced in $\cos\theta$ over the range $-1 \leq \cos \theta \leq 1$, where the angle $\theta$ is taken between the line of sight and the symmetry axis as defined in Equation \ref{eq:theta_bins}. 
Specifically, simulations in our database cover all observing angles with a resolution ranging from $\Delta\theta=\arcsin{(2/54)}\simeq2.1^\circ$ at the equator, to $\Delta\theta=\arccos{(1-2/54)} \simeq15.6^\circ$ near the axes.
This limiting angular resolution \edited{near the axes} is comparable to the
  angles inferred from long-term radio observations of the gamma ray burst jet afterglow \cite{2017ApJ...848L..27T, Evans_2017, Troja_2017, 2018Natur.561..355M, 2020PhRvL.125n1103B, 2020MNRAS.498.5643T}.

\texttt{SuperNu} Monte Carlo radiative transport results have modest but nonzero Monte Carlo error.  The impact of this Monte
Carlo error can be estimated both by resolution studies as well as by simple smoothness diagnostics (e.g., versus angle
and time).  For example, in a previous study \cite{Ristic22}  we performed Gaussian process interpolation over this same training set,
inferring both an estimate of the light curve and a conservative estimate of its variance.  The Monte Carlo error
inherent in the underlying \texttt{SuperNu} simulations can be seen for example in Figure~6 of Ref.~\cite{Ristic22} as short-angular-scale
roughness on top of the overall smooth trend, \edited{with at most $\sim 3\%$ deviation at the $1\sigma$ level}. Later, in Section~\ref{sec:high_res_theta}, we describe a targeted
resolution study.   Based on these investigations, we expect that the \emph{statistical} error of the underlying \texttt{SuperNu} simulations is smaller
than the systematic errors introduced by interpolating between these simulations as  described below.

\subsection{Training Data Generation}
\label{sec:trainingset}

Below we describe the approach taken to generate the simulation library in Ref.~\cite{Ristic22}, hereafter R22. Our training library of $22248$ kilonova light-curve simulations was constructed using iterative simulation placement guided by Gaussian process variance minimization. New simulations were placed with parameter combinations that were identified as having the largest bolometric luminosity variance by our Gaussian process regression approach. In other words, we placed new simulations in regions of parameter space where our bolometric luminosity interpolation root-mean-square uncertainty was largest. The Gaussian process variance $s(\vec{x})^2$ is defined as

\begin{equation}
s(\vec{x})^2 = k(\vec{x}, \vec{x}) - k(\vec{x}, \vec{x}_a)k(\vec{x}_a, \vec{x}_{a'})^{-1}_{aa'}k(\vec{x}_{a'}, \vec{x}) \,,
\label{eq:rms}
\end{equation}
where $\vec{x}$ is the vector of input parameters, $\vec{x}_a$ is the training data vector, the function $k(\vec{x}, \vec{x}')$ is the kernel of the Gaussian process, and the indices $a, a'$ are used to calculate the covariance between inputs $\vec{x}$ and training data $\vec{x}_{a}, \vec{x}_{a'}$ such that if $a = a'$, the variance is 0. In building the simulation library, we only considered the four-dimensional space of ejecta parameters $\vec{x} = [M_{\rm{d}}, v_{\rm{d}}, M_{\rm{w}}, v_{\rm{w}}]$. \edited{Each ejecta parameter combination yields simulations calculated for 54 equally-spaced viewing angles; as such, our training set of 22248 light curves corresponds to a core set of 412 unique ejecta parameter combinations.} 

For this work, we use the aforementioned light curves in the original simulation library as our training set. The light curves used in this work have the same parameters as those used for our light-curve interpolation approach in R22. No additional simulations were produced for the purposes of this work; all training data came from the simulation library presented in R22.

The original training data library consists of $22248$ total light-curve simulations calculated at $264$ times and $54$ angular bins each. We do not perform any coordinate transformations, but rather interpolate directly in our ejecta parameter space and angle.

\subsection{Data Processing}
\label{sec:preprocessing}
The entirety of our 22248 simulations is represented by a tensor, $M_{abc}$, containing the AB magnitudes across the bands described in Section~\ref{sec:sim_setup}, which correspond to a set of input parameters $\vec{x}$. $M_{abc}$ has dimensions of $412 \times 264 \times 54$, corresponding to 412 simulations evaluated at $264$ log-spaced times between $0.125$ and $37.24$ days for $54$ viewing angles equally spaced in $\cos\theta$ for $\theta$ ranging from $0$ to $180^\circ$. \edited{We do not perform any normalization of our inputs or outputs, with ejecta parameters ranging from $-3 \leq \log m / M_\odot \leq -1$ and $0.05 \leq v/c \leq 0.3$ and light-curves ranging from -18 to 8 AB magnitudes.}

We split our four-dimensional ejecta parameter vector $\vec{x}$ into training, validation, and test sets. We use 60\% of the data for the training set, which contains information that the \edited {neural network} uses to learn. Of the remaining 40\%, 20\% is used for the validation check, which tracks how well the \edited{network} generalizes to off-sample inputs during training, and 20\% goes into the test set, which is used to evaluate the \edited{network's} predictions compared to known simulation data. \editprr{The samples for each set are randomly drawn from $\vec{x}$ according to a uniform distribution.} Neither the validation set nor the test set data is used by the \edited{neural network} for learning; therefore, we only use $\sim 247$ simulations for training, while the rest are used in various steps of verifying generalization (i.e. avoiding overfitting to the training data). 

After splitting the data into training, validation, and test sets, we incorporate viewing angle as a fifth input parameter. As mentioned above, the viewing angles in our simulations are equally-spaced in $\cos\theta$ space across 54 angular bins, as presented in the following equation:
\begin{equation}
\theta = \arccos\left(1 - \frac{2(i - 1)}{54}\right) \text{ for i in 1, 2, \dots, 54}
\label{eq:theta_bins}
\end{equation}

Temporarily ignoring training, validation, or test sets, our training library consists of a total of $22248 \times 5$ inputs, as illustrated in the following schematic matrix: 
\begin{center}
$\vec{x}_{\theta}$ = $\begin{bmatrix}
m_{d, 1} & v_{d, 1} & m_{w, 1} & v_{w, 1} & \theta_1 \\
m_{d, 1} & v_{d, 1} & m_{w, 1} & v_{w, 1} & \theta_2 \\
m_{d, 1} & v_{d, 1} & m_{w, 1} & v_{w, 1} & \theta_3 \\
\vdots & \vdots & \vdots & \vdots & \vdots \\
m_{d, 1} & v_{d, 1} & m_{w, 1} & v_{w, 1} & \theta_{54} \\
m_{d, 2} & v_{d, 2} & m_{w, 2} & v_{w, 2} & \theta_1 \\
m_{d, 2} & v_{d, 2} & m_{w, 2} & v_{w, 2} & \theta_2 \\
\vdots & \vdots & \vdots & \vdots & \vdots \\
m_{d, 412} & v_{d, 412} & m_{w, 412} & v_{w, 412} & \theta_{54} \\
\end{bmatrix}$
\end{center}

\subsection{\edited{Neural Network} Architecture and Training}
 \begin{figure*}
    \centering
    \includegraphics[width=\linewidth]{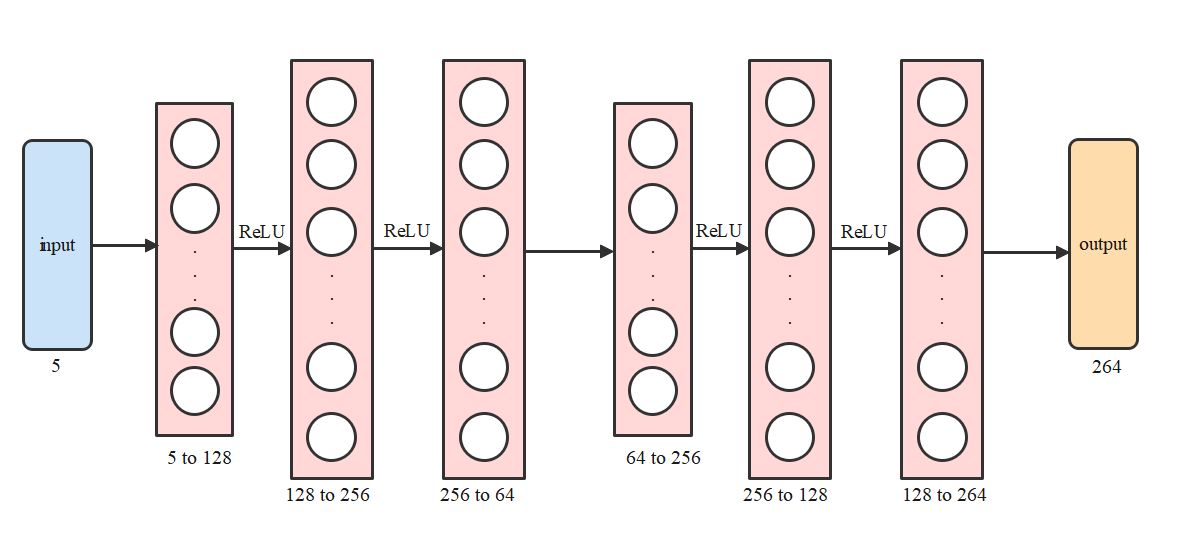}
    \caption{A visual representation of the \edited{neural network} architecture. The blue block represents our five-dimensional inputs $\vec{x}_{\theta}$. The pink blocks represent hidden layers, with labels under each block representing the input and output dimensions of each. Unlabeled right arrows indicate a linear mapping between layers, while those labeled ``ReLU" have a Rectified Linear Unit activation function applied to their outputs. The orange block represents the \edited{neural network} prediction in the form of a $264 \times 1$ vector matching the length of our broadband light curves.}
    \label{fig:architecture}
 \end{figure*}

We use a \edited{standard feed-forward} neural network called \edited{a multi-layer perceptron (MLP)}. Figure~\ref{fig:architecture} shows our \edited{MLP} architecture. The input data of dimension 5 (blue block) \edited{is propagated through the hidden layers of the MLP (pink blocks). These hidden layers} apply a sequence of linear and non-linear transformations (black arrows) to progressively map the input to a higher-dimensional space. \edited{The network} has six fully-connected layers (pink blocks) \edited{of dimension 128, 256, 64, 256, 128, and 264, respectively,} which are followed by Rectified Linear Unit (ReLU) activation functions, \edited{except for the middle two layers}. The final layer reduces the dimension to a $264 \times 1$ vector, which matches the length of our light curves. 

\edited{For each observing band, we train a separate neural network and compare its predictions with} the simulation results. During training, the \edited{neural network's} predictions for the inputs in the validation set are compared to the simulation data for those same inputs. The residual, or difference, between the two is evaluated by the mean squared error (MSE) loss function, which we use to measure the average squared difference between simulation data and the \edited{neural network} prediction. We calculate the MSE according to
\begin{equation}
    \label{eq:mse}
    MSE = \frac{1}{264} \sum_{i=1}^{264}(y_i - \hat{y_i})^2
\end{equation}
where $y_i$ is the corresponding simulation data at time $i$ and $\hat{y_i}$ is the predicted value from the \edited{MLP} model at the same time. We show the evolution of the training and validation losses in Figure~\ref{fig:loss} for the \emph{g}-band \edited{network}. 

We train a separate \edited{neural network} for each of the broadband filters described in Section~\ref{sec:sim_setup}. 
Each \edited{network} is trained for 1000 epochs, as this is enough time for the validation loss to convincingly stabilize at a floor value without beginning to increase, indicating overfitting. \editprr{We use an Adam optimizer and train each network in batches with a batch size of 32. Our initial learning rate is $2~\times~10^{-4}$ with a decay rate of 5\% every 10 epochs.}

\subsection{Neural Network Performance}
\label{sec:ae_peformance}

\begin{figure}
    \centering
    \includegraphics[scale=0.55]{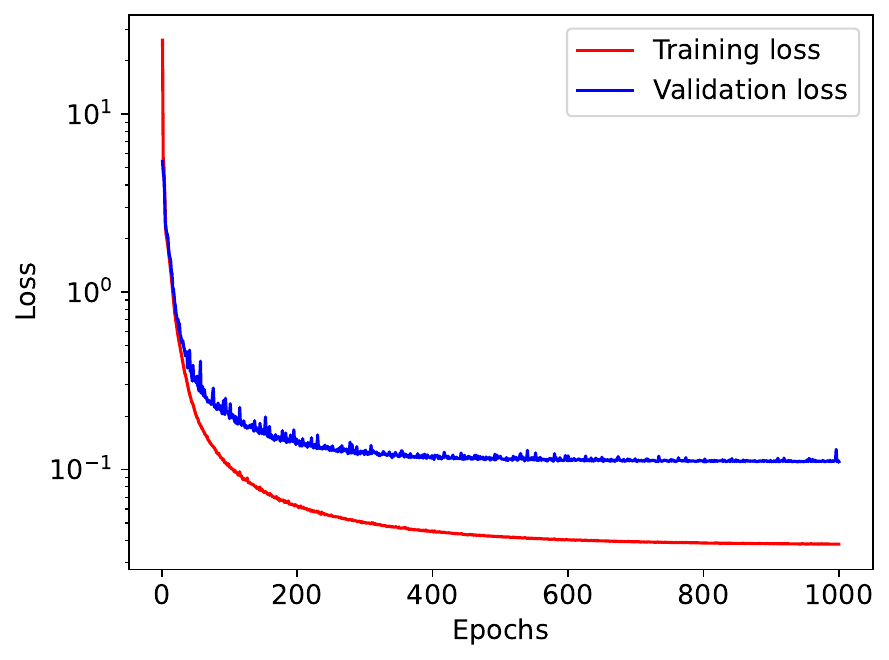}
    \caption{Training (red) and validation (blue) loss curves as a function of training epoch for the \emph{g} band. The loss values reported here are mean squared error (MSE) as defined in Equation \ref{eq:mse}. Decreasing values of loss indicate better agreement between the model and the training data. Our training and validation loss decreases over the course of the 1000 epoch training period. The validation loss appears to converge at around 500 epochs, although we continue to 1000 epochs as we do not implement an explicit convergence criterion.}
    \label{fig:loss}
 \end{figure}

\begin{figure*}
    \centering
    \includegraphics[width=0.48\textwidth]{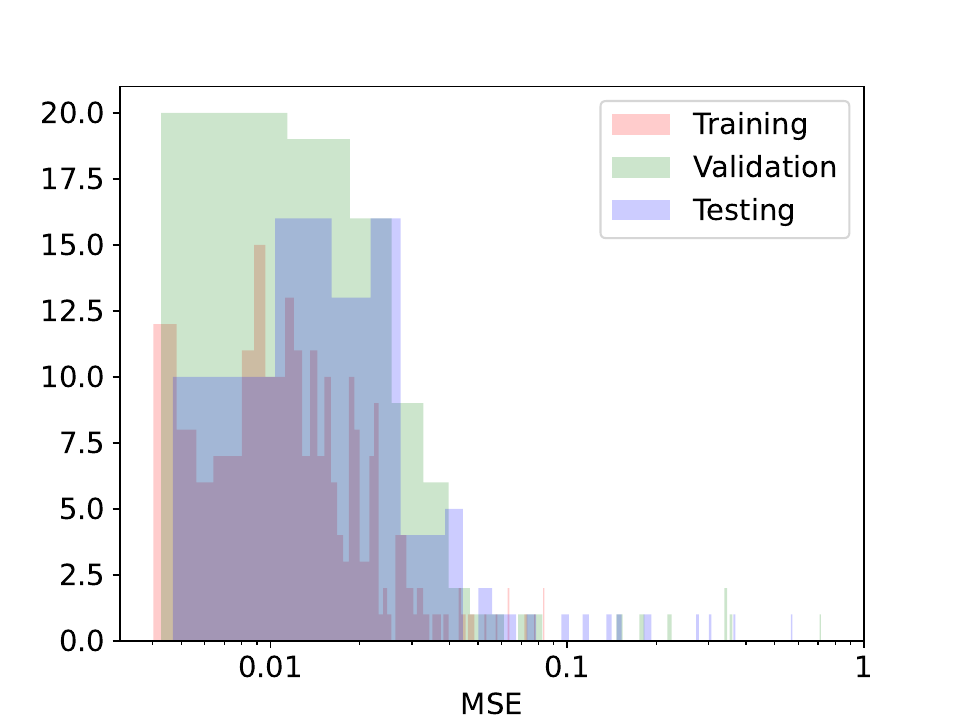}
    \includegraphics[width=0.48\textwidth]{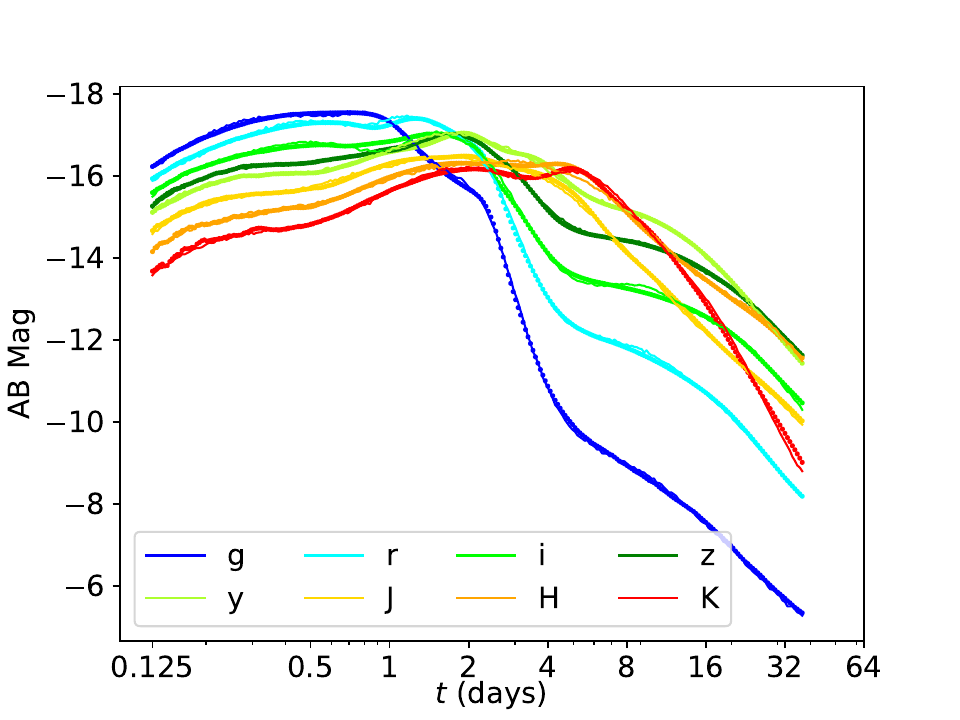}
    \includegraphics[width=0.48\textwidth]{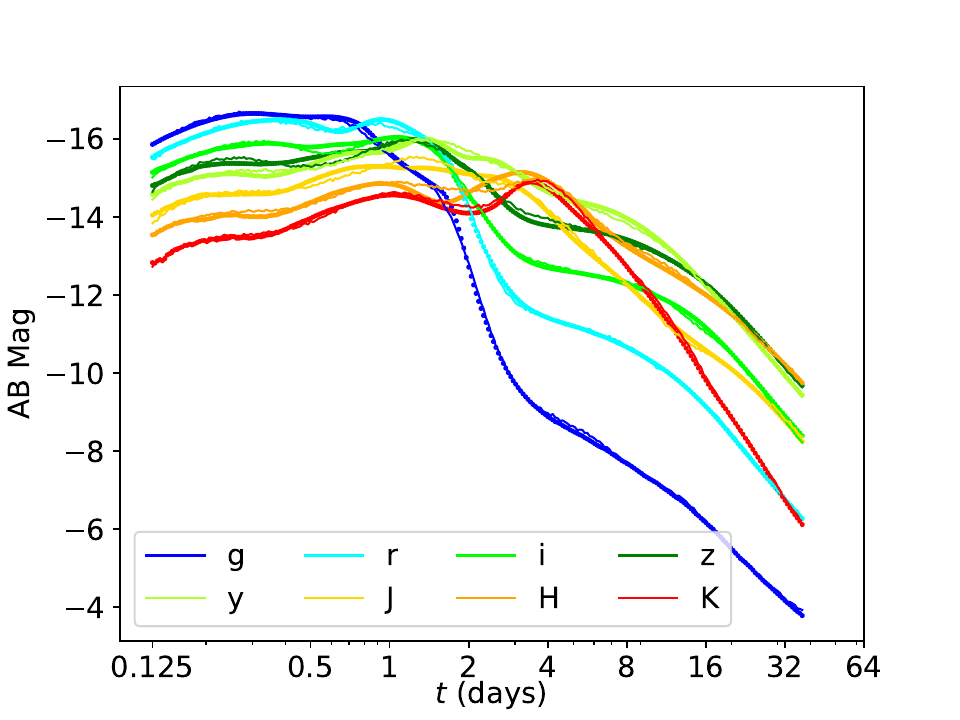}
    \includegraphics[width=0.48\textwidth]{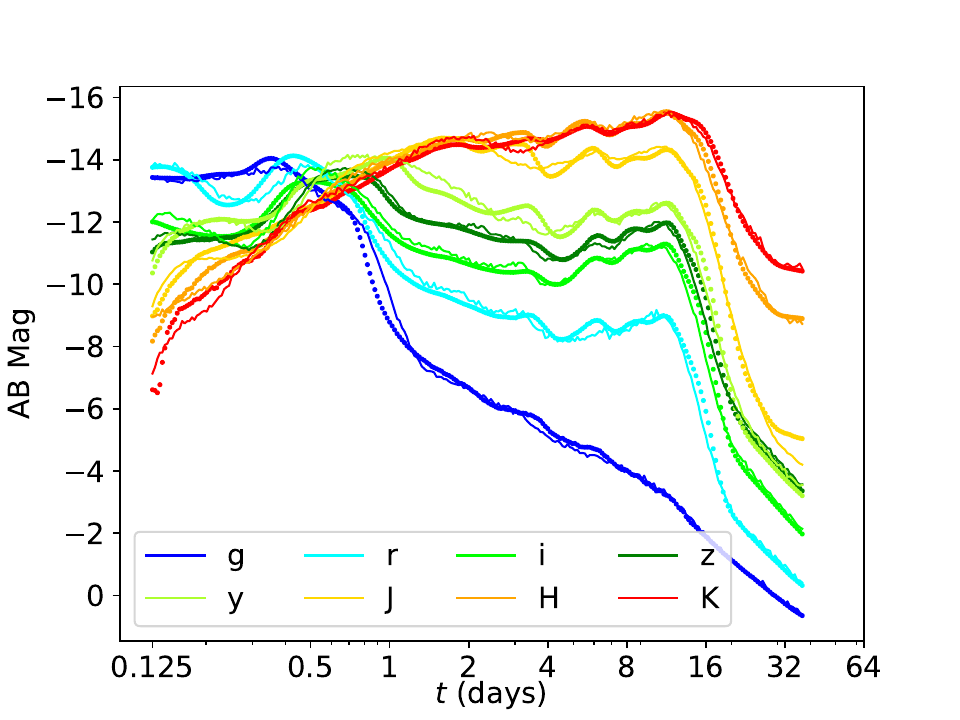}
    \caption{\edited{\emph{Top left}: A histogram of MSE values, averaged across all bands by the number of observations, which characterize the deviation of the MLP's predictions from the true simulation library light curves. For simplicity, we assume a fixed viewing angle of $\theta = 0^\circ$ for each light curve and evaluate the MSE only for the 412 unique ejecta parameter combinations. \emph{Top right}: True (points) and predicted (lines) light curves for a randomly drawn set of ejecta parameters \editprr{(still assuming $\theta = 0^\circ$)} with $MSE < 0.01$. \emph{Bottom left}: Same as top right, but for $0.01 \leq MSE \leq 0.1$. \emph{Bottom right}: Same as top right, but for $MSE > 0.1$.}}
    \label{fig:testset}
\end{figure*}

Figure~\ref{fig:loss} indicates that our training loss undershoots the validation loss starting at $\sim 100$ epochs, but the
nearly perfect recovery of the test-set light curves shown in Figure~\ref{fig:testset} shows no indication of
overfitting to the training data. For inference applications, our \edited{neural networks} can generate the outcomes corresponding to
five million five-dimensional inputs  in about one minute.
Training each \edited{neural network} takes 20 minutes on a 2022-edition Macbook Pro with an M2 chip \edited{using the CPU}. Since all the bands are
independent and can be trained simultaneously in parallel, training all the emulators can be completed in this same 20
minute interval.  Our training time is half the value reported in Ref.~\cite{2021arXiv211215470A}, though it is unclear
whether their reported time assumes training in parallel or in serial. 

\edited{The top left plot in Figure \ref{fig:testset} shows a histogram of the MSE values when evaluating the simulation library parameters using the neural network. We evaluate only the 412 unique ejecta parameter combinations, fixing the viewing angle to $\theta = 0^\circ$ in each case. In containing simulations from the training, validation, and test sets, this histogram represents the neural network's on- and off-sample fidelity. The light-curve plots in Figure \ref{fig:testset} show random draws from the MSE histogram for $MSE < 0.01$ (top right), $0.01 \leq MSE \leq 0.1$ (bottom left), and $MSE > 0.1$ (bottom right)}.

We note that, as with the emulators presented in Ref.~\cite{Ristic22}, predictions for inputs with low-mass ($\log(M) \sim -3$) components or viewing angles $\theta \sim 90^\circ$ may deviate substantially from expectations. In \edited{Monte Carlo} radiative transfer simulations of a kilonova, the representation of very low-mass components and viewing angles in the plane of the binary present a significant challenge. In order to study the dynamics of the energy distribution, a finite number of particles are employed to represent photons escaping from the system, forming ``packets" of energy. \editprr{It is well known that at low Monte Carlo particle counts, tallies of escaping flux can even have systematic error \cite{2018ApJ...861...80C}. In our simulations, we mitigate this effect with implicit capture variance reduction and discrete diffusion Monte Carlo \cite{SuperNu}. However, given a fixed number of source particles per time step, we use a metric for allocating particles per spatial cell that is sub-linearly proportional to emissivity, so that the statistical quality of a spatial cell in one ejecta component is affected by the other ejecta component.} Consequently, when dealing with extremely low-mass components or viewing angles that look into the high-opacity dynamical ejecta, the simulations become particularly sensitive to Poisson noise due to reduced photon count. \edited{This effect, particularly with respect to low-mass ejecta, likely arises from our \texttt{SuperNu} simulations preferably sampling photon packets from higher-energy regions of the ejecta.} In future studies, we hope to enhance the simulation interpretation under these conditions by way of an increase in photon packet count.

\section{Bayesian inferences  with the neural interpolator}
\label{sec:results}

\subsection{Parameter Inference Methodology}
\label{sec:pe_methods}

As in R22, we infer the parameters of the kilonova AT2017gfo using our interpolated light curves and the AT2017gfo
photometric data. The AT2017gfo data is originally presented in \cite{2017PASA...34...69A, 2017Natur.551...64A, 2017Sci...358.1556C, 2017ApJ...848L..17C, Diaz17, 2017Sci...358.1570D, Evans_2017, 2017SciBu..62.1433H, 2017Sci...358.1559K, 2017ApJ...850L...1L, 2017Natur.551...67P, 2018ApJ...852L..30P, shappee2017early, 2017Natur.551...75S, 2017ApJ...848L..27T, Troja_2017, 2017PASJ...69..101U}.
We use the RIFT framework \cite{Wofford2023} to \edited{adaptively perform the  Monte Carlo
  integral  and generate samples} using a reduced $\chi^2$ statistic. The parameter priors are the same as in R22, with uniform ejecta parameter priors of $-3 \leq \log m / M_\odot \leq -1$ and $0.05 \leq v/c \leq 0.3$ and a Gaussian angle prior with $\mu = 20$ and $\sigma = 5$ degrees. Unlike before, in this work we employ an adaptive volume Monte Carlo integrator, following closely the approach outlined in Ref. \cite{2023PhRvD.108b3001T}. The adaptive volume integrator allows for more efficient sampling given the higher-dimensional space being explored in this work.

Each sample $\vec{x}_{\theta}$ is evaluated by the \edited{MLP} to produce a light-curve prediction $\hat{y}$ for every one of the grizy JHK broadband filters. We calculate the residual between the \edited{MLP} prediction $\hat{y}$ and the AT2017gfo observed data $d$ for every band $B$ by way of the reduced-$\chi^2$ statistic

\begin{equation}
    \chi^2 = \sum_{B, i} \frac{(\hat{y}_{i, B} - d_{i, B})^2}{\sigma_i^2 + \sigma_{\rm sys}^2}.
\end{equation}

In our $\chi^2$ residual calculation, we include observational uncertainties from the AT2017gfo data $\sigma_d$, as well
as systematic uncertainties $\sigma_{\rm{sys}}$ which we use as a catch-all term to encompass all uncertainties,
quantifiable or otherwise, associated with the \edited{neural network} interpolation process.  As outlined above and as discussed in
greater quantitative detail  in Section~\ref{sec:high_res_theta}, we adopt a systematic modeling uncertainty,
$\sigma_{\rm{sys}}$, of 0.5 magnitudes for our inference analysis. 

\edited{For inference, we  adopt a purely Gaussian likelihood based on these residuals}
\begin{equation}
\ln {\cal L} = -\frac{\chi^2}{2} - \frac{1}{2}\ln (2\pi)^N\sum_k(\sigma_d^2+\sigma_{\rm sys}^2)
\end{equation}
\edited{where $N$ is the number of observations.  As our inference is performed via adaptive Monte Carlo integration,
  the reliability of our posterior can be expressed in terms of a number of effective samples $n_{\rm eff}$.  [Several
  different conventions exist for this number; see the appendix of \cite{gwastro-RIFT-Update} for discussion.]   For this study,
  we terminate our analyses when $n_{\rm eff} \simeq 10^3$. }

\begin{figure}
    \centering
    \includegraphics[width=\linewidth]{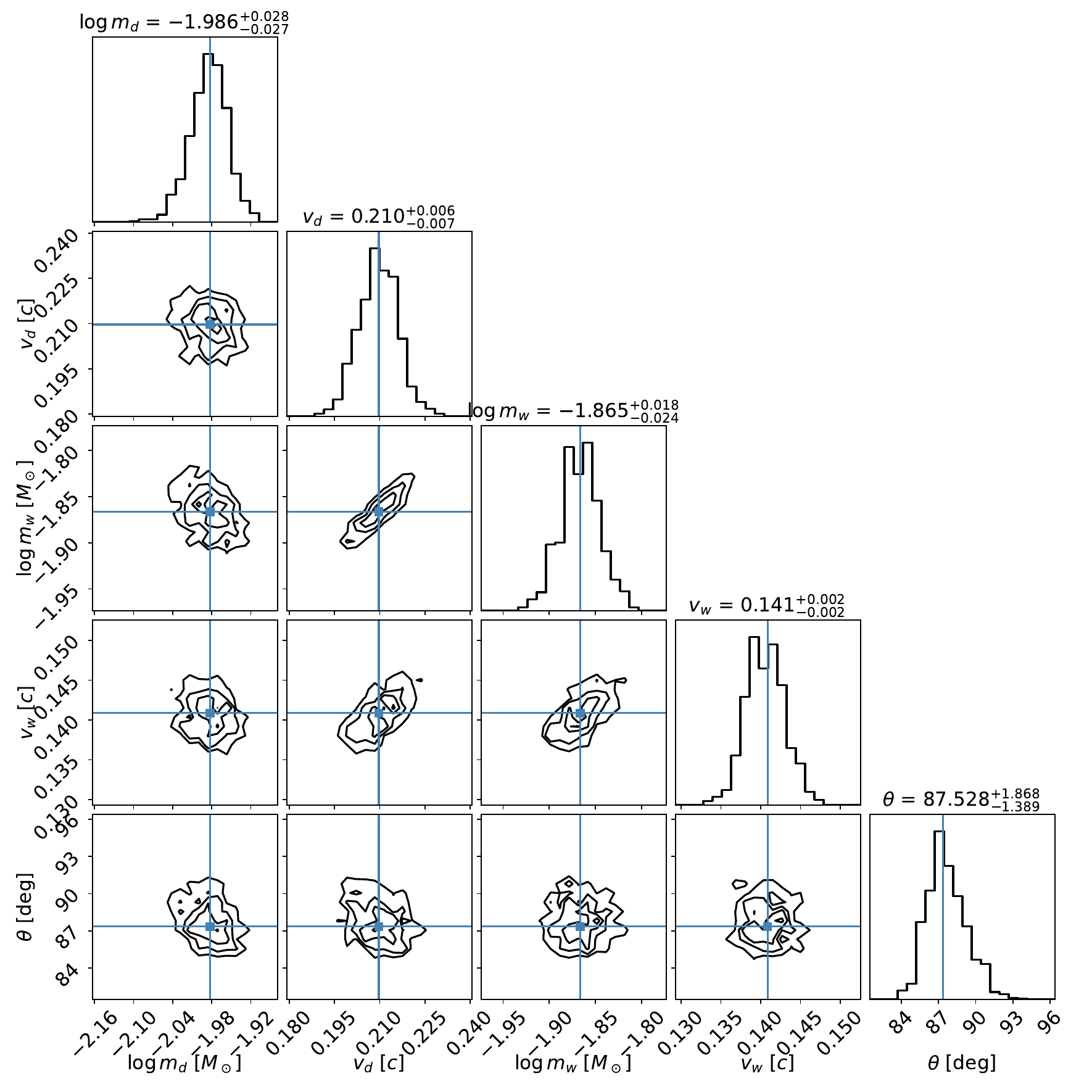}
    \caption{\edited{Posteriors derived from a single randomly-generated synthetic kilonova source are consistent with its assumed
        model parameters.  For this analysis, we have adopted the
        ``zero noise'' realization, where the kilonova light curve is precisely equal to its expected value. }}
    \label{fig:single_injection_posteriors}
\end{figure}

\edited{To validate our inference strategy, we constructed random synthetic sources, with kilonova model parameters drawn from our prior (albeit
  adopting a uniform rather than gaussian angular prior) and using observational times and uncertainties precisely matching the
  AT2017gfo cadence and instruments.  For our synthetic sources, the expected light curve is generated using our neural
  network.   As demonstrated with one example in Figure \ref{fig:single_injection_posteriors}, our inferences are always consistent with the injected kilonova
  parameters.  To demonstrate that our implementation retains statistical purity, we also used 100 random synthetic sources
  to perform a conventional probability-probability (PP) test, available in an Appendix. }

\edited{While we  employ the fixed $\sigma_{\rm sys}=0.5$ for most of our studies, in order to validate our results we also perform a
  few selected analyses with different choices, on the one hand adopting different discrete choices and on the other
  treating $\sigma_{\rm sys}$ as a continuous unknown model parameter.}

\edited{The analysis in Figure \ref{fig:single_injection_posteriors} demonstrates that, if the underlying model is
  correct, comparison with AT2017gfo-like observations should very tightly constrain each of this model's parameters.  This fiducial result thus
  has qualitatively different behavior than our and others' prior analyses of AT2017gfo, where posterior inferences
  arrive at much broader  posterior intervals, as discussed below.  That said, the posterior shown above is consistent
  with the standard Fisher matrix estimate of the inverse covariance matrix $\Gamma=\Sigma^{-1}$, derived for example by taking  (the expected value over
  noise realizations of) a second-order Taylor series expansion of the log-likelihood as $\ln {\cal L} = \ln {\cal L}_o -
  2^{-1} \Gamma_{ab} (x-x_*)_a(x-x_*)_b$ where $x_*$ are the true synthetic parameters:}
\begin{align}
\Gamma_{ab} = \sum_{B,i}   
\frac{1}{\sigma_{i}^2+\sigma_{\rm sys}^2} 
  \frac{\partial \hat{y}_B }{\partial   x_a }
 \frac{\partial \hat{y}_B}{\partial x_b}_{x=x_*,t=t_i}
\end{align}
\edited{In this expression, only first-order derivatives appear because we assume that the model has no systematic bias such
  that $\E{d}=\hat{y}$; however, this simple estimate also arises inevitably using the large-amplitude ``linear signal
  approximation'' \cite{2008PhRvD..77d2001V}.
This  Fisher matrix can be estimated to order of magnitude by replacing the derivatives $\partial \hat{y}_B/\partial x_a$
by the ratio $\Delta y_B/\Delta x_a $, which for the mass parameters we approximate as $2/2=1$, so the Fisher matrix is
approximately $\Gamma\simeq N/\sigma_{\rm sys}^2$ and the posterior  in each mass hyperparameter should have a
one-standard-deviation scale of order
$1/\sqrt{\Gamma}\simeq \sigma_{\rm sys}/\sqrt{N} \simeq 0.5/\sqrt{333} \simeq 0.027$.
}

\subsection{AT2017gfo Parameter Inference}
\label{sec:inference}

\begin{figure}[ht!]
    \centering
    \includegraphics[width=\columnwidth]{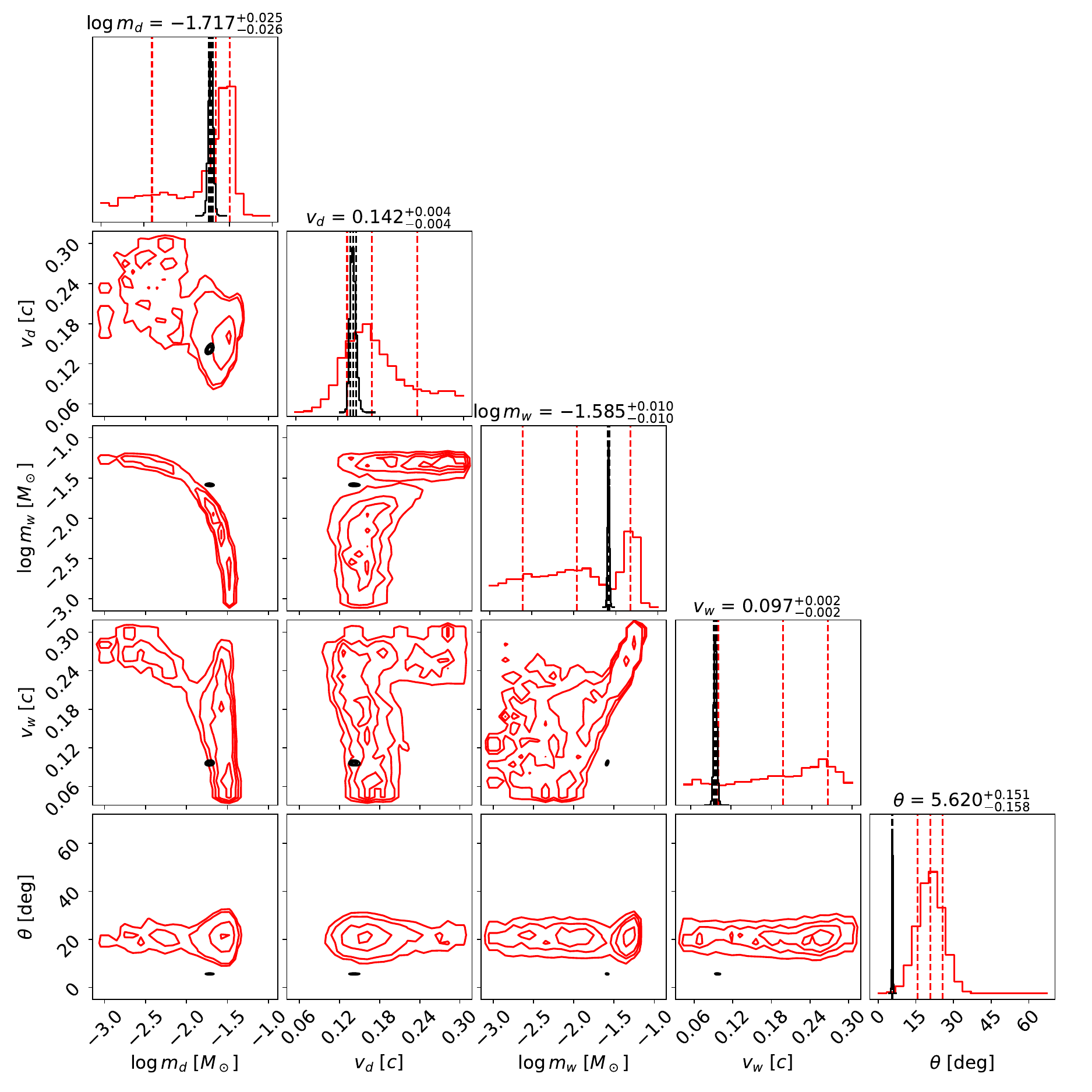}
    \caption{In black are shown the posterior distributions for the ejecta parameters and viewing angle that most closely reconstruct the AT2017gfo observational data using the \edited{MLP}. In red, we overlay the posterior distributions for the same parameters when using predictions generated by the Gaussian process presented in R22. The values at the top of each column represent the median posterior values for the inference performed in this paper using the \edited{MLP}.}
    \label{fig:posteriors}
\end{figure}

\begin{figure}[ht!]
    \centering
    \includegraphics[width=\columnwidth]{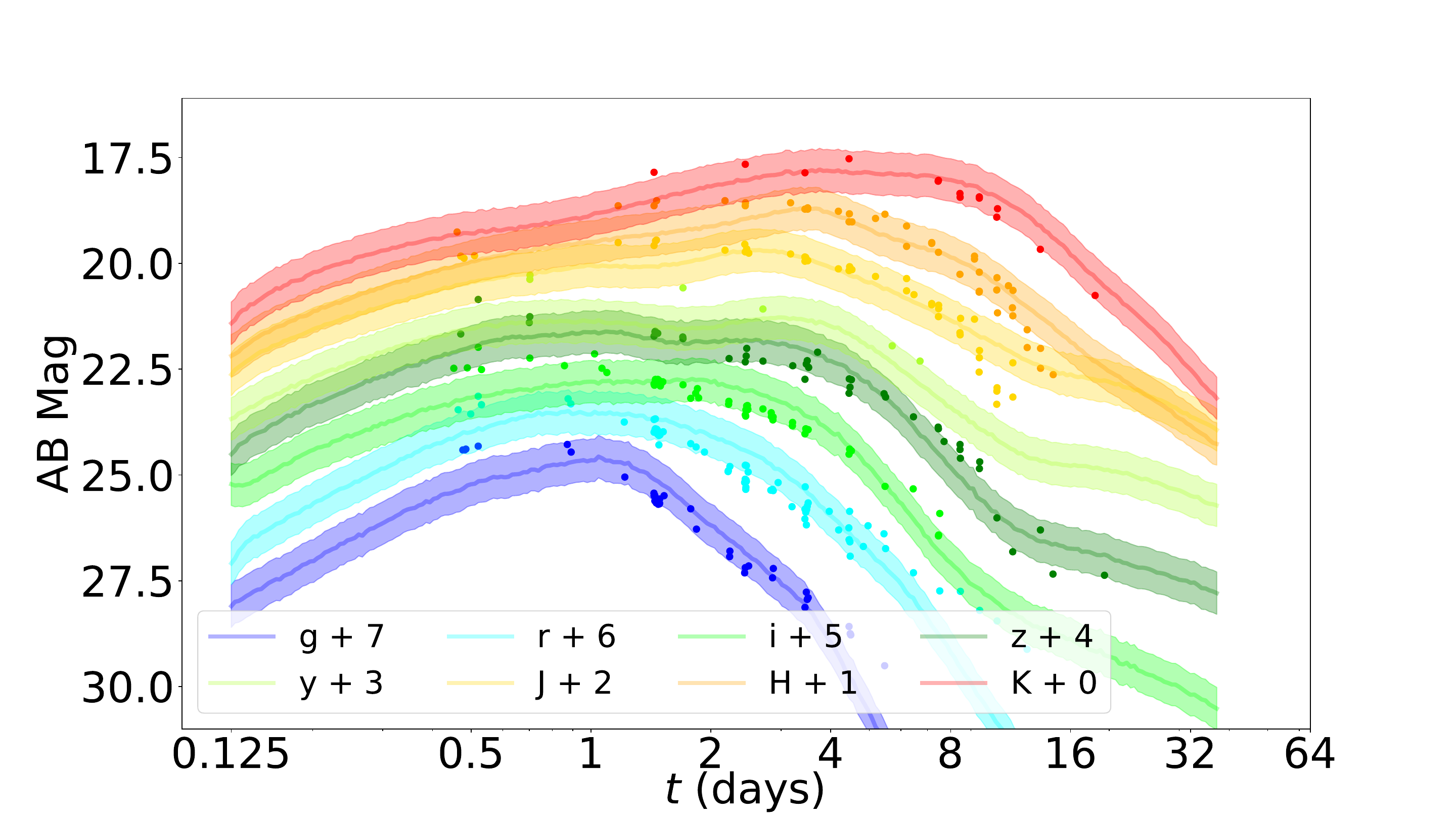}
    \caption{Light curves generated by the \edited{MLP} for the median parameters presented in Figure \ref{fig:posteriors} (lines with shaded regions) with AT2017gfo observational data overplotted (scatter points). Despite the different values recovered between this analysis and that of R22, especially for the wind ejecta parameters and $\theta$, the \edited{MLP} prediction is able to replicate the observations surprisingly well.}
    \label{fig:lc}
\end{figure}

The posterior distributions for our input parameters $\vec{x}_{\theta}$ are plotted in Figure~\ref{fig:posteriors}. The black posteriors represent the parameters identified in this study. As a direct comparison to the results of R22, we overplot the posterior distributions from that study in red. Despite using the same training data and comparing to the same observational data, the Gaussian process emulator and the \edited{MLP} emulator recover posterior distributions with substantially different median values and distribution widths; \editprr{however, we note that in R22, each time step is fit individually, whereas in this work we fit the entire light curve for a given set of input parameters}. The dynamical ejecta parameters $\log m_d$ and $v_d$ are similar between the two emulators, but the wind ejecta parameters $\log m_w$ and $v_d$, as well as the viewing angle $\theta$, are quite different, with two of the three parameters inferred by the \edited{MLP} residing outside of the GP inference 1$\sigma$ limits. 

We verify the fidelity of the \edited{MLP} inference results by generating light curves corresponding to the parameter values identified at the top of each column in Figure~\ref{fig:posteriors}. These light curves are shown in Figure~\ref{fig:lc} and indicate that, assuming a 0.5 magnitude systematic uncertainty, the inferred \edited{MLP} parameters do indeed replicate the AT2017gfo data to a reasonable degree of accuracy. While the majority of data is well replicated, the early-time \emph{g} and \emph{r} bands and the late-time \emph{J} and \emph{H} bands deviate slightly outside of our uncertainty bands.

The recovery of $\theta \approx 6^\circ$ is surprising for several reasons. First, the recovered angle was modestly
offset from the Gaussian adopted as our inclination prior. \edited{Although different studies find a variety of viewing angles associated with AT2017gfo \cite{2017ApJ...848L..27T, Evans_2017, Troja_2017, 2018Natur.561..355M, 2018ApJ...860L...2F, 2020PhRvL.125n1103B, 2020MNRAS.498.5643T}, none indicate that the viewing angle is as low as our inference suggests.} Second, and more
importantly, for angular binning described by Equation~\ref{eq:theta_bins}, the first angular bin encompasses all
emitted photons for viewing angles $\sim 0-16^\circ$. Therefore, by recovering a narrowly-peaked posterior around
$\theta \approx 6^\circ$, the \edited{MLP} seems to indicate that it can identify angular variations within a single angular bin
at a resolution much finer than what is provided by training data. In other words, these inference results are either
overly constrained, or the \edited{MLP} is able to identify fine angular variations in the light curves when trained on radiative
transfer simulations using a coarser angular grid. 

\begin{figure}
    \centering
    \includegraphics[width=\columnwidth]{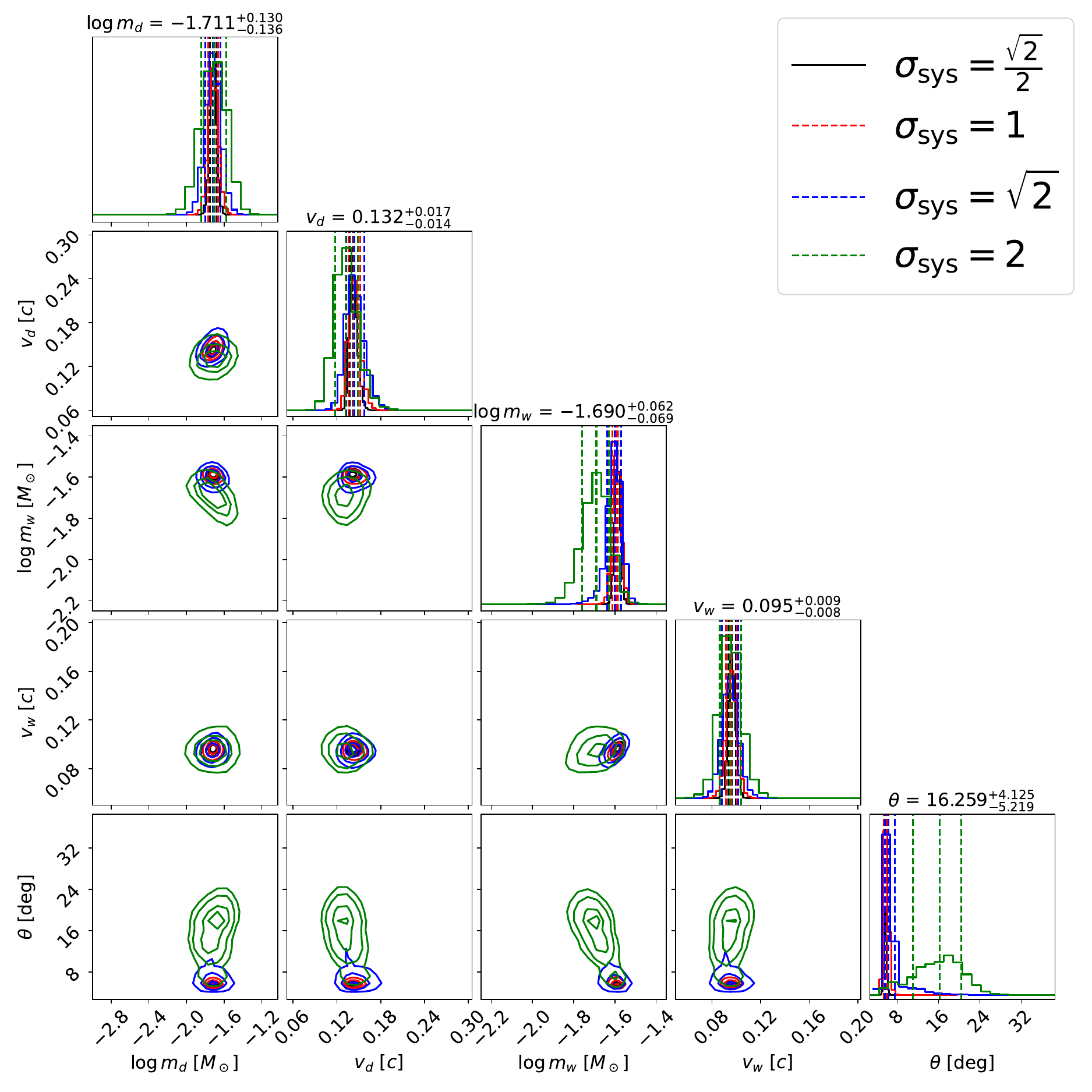}
    \includegraphics[width=\columnwidth]{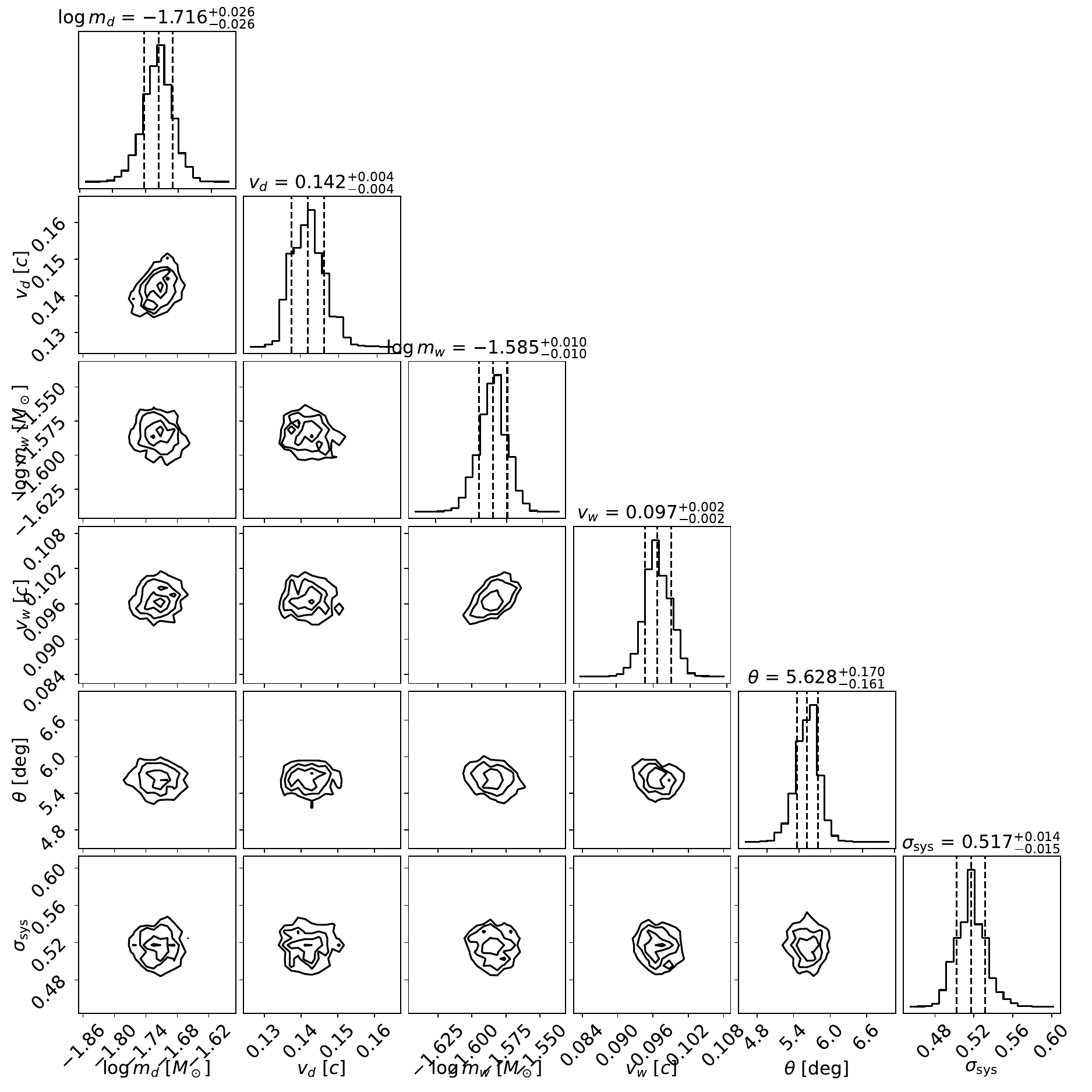}
    \caption{\emph{Top panel}: \edited{Posterior} distributions using successively larger values of
      $\sigma_{\rm{sys}}$ during inference. Recovery of angles closer to the angle prior of $20^\circ$ begins with
      $\sigma_{\rm{sys}}$ = 2, suggesting underestimation of systematic error..
\edited{\emph{Bottom panel}: Posterior distributions inferred if $\sigma_{\rm sys}$ is treated as an unknown parameter,
  a priori uniformly distributed.}
}
    \label{fig:sigma_sys}
\end{figure}

\edited{One conceivable explanation for the narrow posterior distribution seen in Figure \ref{fig:posteriors} is an underestimate of the
  underlying systematic error.  To investigate this possibility,
 Figure~\ref{fig:sigma_sys} shows the results of inferences performed when adopting dfferent choices for 
the white-noise systematic error parameter $\sigma_{\rm sys}$, adopting both discrete and continuously-distributed
choices for this parameter.   In all scenarios, we infer similar \editprr{ejecta} parameters for AT2017gfo, even though we allow for
several magnitudes of potential systematic uncertainty.   Conversely, this direct comparison between our models and the
data directly infers a value for our systematic uncertainty parameter consistent with our fiducial choice. \editprr{As such, we conclude that our inference about the ejecta parameters remains robust for all analyses, with the viewing angle inference susceptible to systematic uncertainty assumptions. However, we approximate the light curves to deviate by $\sim 0.35$ magnitudes between $0^\circ$ and $45^\circ$ viewing angles \cite{RisticThesis}. This deviation for the bolometric light curve still falls below our assumed systematic uncertainty using broadband light curves, and thus remains consistent with the baseline analysis.} }

\subsection{Investigating \edited{MLP Predictions} Near Inferred AT2017gfo Parameters}
\label{sec:high_res_theta}

\begin{figure}
    \centering
    \includegraphics[width=\columnwidth]{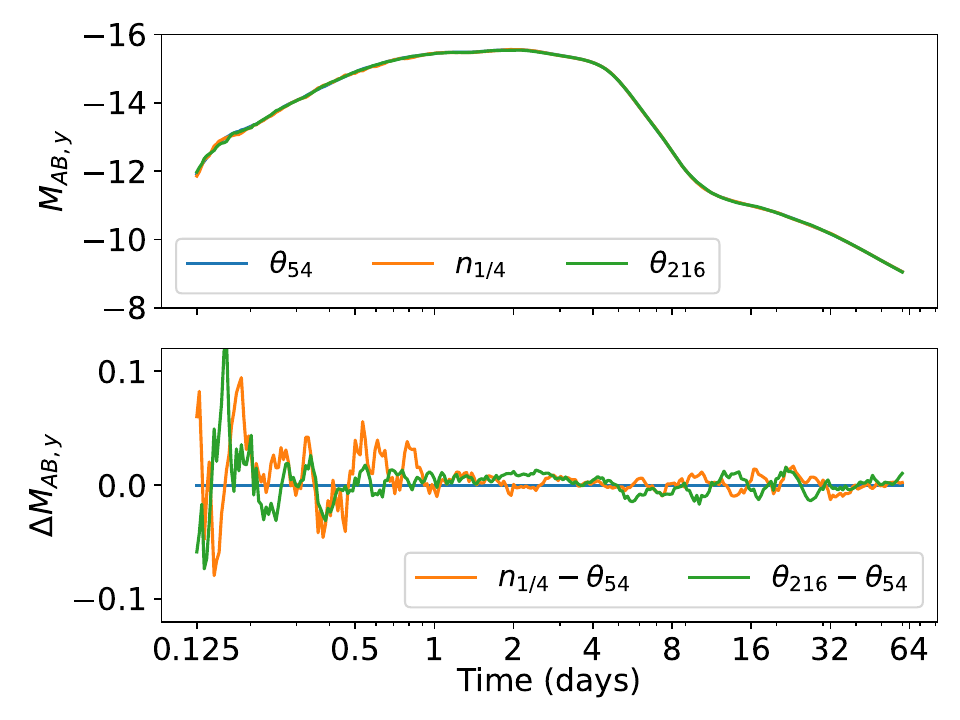}
    \caption{Plots of \texttt{SuperNu} \emph{y}-band light curves for a simulation like the ones described in Section~\ref{sec:sim_setup} ($\theta_{54}$), a simulation like $\theta_{54}$, but with one-quarter as many photon packets ($n_{1/4}$), and a simulation like $\theta_{54}$, but with four times greater angular resolution ($\theta_{216}$). The top panel indicates that, on a macroscopic scale, the simulations are identical. The bottom plot indicates that deviations do exist, likely attributed to statistical noise from the increase in angular bins, or matching reduction in packet count. The ejecta parameters used to create these simulations are those presented in Figure~\ref{fig:posteriors}.}
    \label{fig:noise}
\end{figure}

The extremely narrow posterior distribution in kilonova parameters and angle motivates a focused investigation of our
training simulations and \edited{MLP} model in the neighborhood of that posterior.  
As a first step, we performed followup \texttt{SuperNu} simulations at the inferred parameters using a higher angular resolution.  Specifically, we increased the resolution by a factor of four to get a total of 216 angular bins. By reducing the number of photon packets in each angular bin by a factor of four, we also increase statistical noise by a factor of two. 
To ensure that our finer angular resolution analysis is not affected by this increase in statistical noise, we compare three separate simulations in Figure~\ref{fig:noise}. The blue curve, labeled $\theta_{54}$, shows a \texttt{SuperNu} simulation using the parameters from Figure~\ref{fig:posteriors}, hereafter $x_{\rm{MLP}}$ and a standard 54-bin angular grid, as used in the training data simulations. The orange curve, labeled $n_{1/4}$, shows a \texttt{SuperNu} simulation with the same exact parameters as $\theta_{54}$, except it uses one-quarter as many photon packets in the simulation. If the statistical noise described above were significant, the noise in $n_{1/4}$ should be much more pronounced than in $\theta_{54}$. Finally, the green curve, labeled $\theta_{216}$, shows a \texttt{SuperNu} simulation using the same parameters as $\theta_{54}$, but with a 216-bin angular grid, representing a factor of four increase in resolution. All three light curves show AB magnitude in the \emph{y}-band as a function of time in days. 
As seen in Figure \ref{fig:noise}, our followup simulations agree with one another, with small Monte Carlo error
comparable to our initial estimate and  small compared to our adopted systematic uncertainty ($\sigma_{\rm sys}=0.5$).

\begin{figure}
    \centering
    \includegraphics[width=\columnwidth]{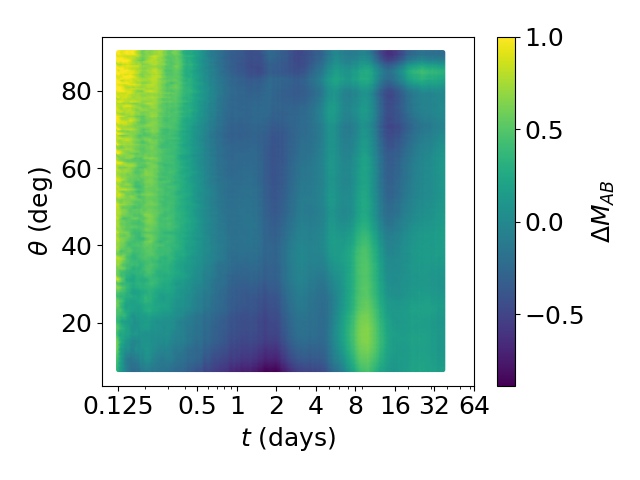}
    \caption{Deviations in \emph{y}-band AB magnitude ($\Delta M_{\rm{AB}}$) between a \texttt{SuperNu} simulation with
      four times the usual angular resolution ($y_{\rm{sim}}$) and the \edited{MLP} prediction for the parameters reported in
      Figure~\ref{fig:posteriors} ($y_{\rm{MLP}}$) as a function of time $t$ and angle $\theta$. 
     Notably, this comparison between our \edited{MLP} (trained at low angular resolution) and the followup simulation (performed
     at high angular resolution) does \emph{not} exhibit strong small-scale variation along the $\theta$ direction
     (i.e., between adjacent angular bins in the original training set and outside the original training resolution), suggesting that the \edited{MLP}
     correctly interpolates to smaller angular scales.
}
    \label{fig:diff}
\end{figure}

We then compare our high angular resolution simulation $\theta_{216}$ to the predictions of the \edited{MLP}, computing the difference between model
and prediction at all times and simulation angles.   Figure~\ref{fig:diff} shows the residual $\Delta M_{\rm{AB}}$ in the \emph{y}-band when we take the absolute difference between $\hat{y}_{\rm{MLP}}$ and $y_{\rm{sim}}$, capped at a maximum difference of 1 magnitude. The residual values $\Delta M_{\rm{AB}}$ are initially evaluated for the 216 discrete angular bins; for visual clarity and diagnostic power, we linearly interpolate $\Delta M_{\rm{AB}}$ across time $t$ and angle $\theta$ using the \texttt{RegularGridInterpolator} function from the \texttt{scipy.interpolate} library \cite{scipy}.
Figure~\ref{fig:diff} exhibits both expected and unexpected behaviors. The large mismatch at early times when photon
count is low, particularly as $\theta$ approaches $90^\circ$ where the higher opacity dynamical ejecta further reduces
photon count, is to be expected. The \edited{MLP} is not fitting the light curves in this region well as there are too few
photons available in the training data. However, the low mismatch (i.e. good fitting) dark blue regions in the plot
indicate some sort of structure in the \edited{MLP's} underlying ability to reproduce light curves at different times and
angles. While we only include the plot of $\Delta M_{\rm{AB}}$ for the \emph{y}-band in Figure \ref{fig:diff}, it is worth
noting that the same behavior can be observed in these plots for all bands; predictable, low photon count systematics are
identified in expected regions, but other, unexpected regions of increased systematic error manifest in different regions of the parameter space. 
Overall, however, the systematic uncertainties illustrated here are consistent with our expectations from the reported
validation loss: a conservative systematic error of $\sqrt{MSE_{val}}\simeq 0.5$ reflects our  overall uncertainty.
For this reason, we adopted this systematic uncertainty in our parameter inferences above. 
This systematic uncertainty is substantially more conservative than the uncertainties adopted in R22.


\subsection{Inference Using Broadband Data Subsets}
\label{sec:blue_vs_red}

\edited{The investigations in Section~\ref{sec:high_res_theta} have introduced a surprise.  On the one hand, our neural network reliably reproduces its training and validation data, including followup off-sample simulations performed at higher resolution.  Though not shown here, we have also confirmed that the neural network agrees well with the surrogate provided in R22, using a small sample of randomly-selected ejecta parameters.  On the other hand, the inferences obtained in Section~\ref{sec:high_res_theta} by comparing all AT2017gfo kilonova observations to our MLP produce strikingly different results than R22. However, as noted in R22 and other works, most investigations have some tension between their models and the data, particularly in the bluer bands.  In this section, motivated by this discrepancy, we also examine the effects on our AT2017gfo parameter inference when we use only specific subsets of the observational data. We perform two additional parameter inference runs using two categories of broadband data subsets: blue bands represented by the \emph{griz} data and red bands represented by the \emph{yJHK} data. 

The posteriors in Figure~\ref{fig:blue_vs_red} show how these band-limited results compare with each other, the results of R22, and the all-band analysis presented in Section~\ref{sec:high_res_theta}. The most apparent result is that the angle prior is recovered in both sets of posteriors, and both cases match the R22 results well. The other interesting feature is that the red \emph{yJHK} posterior matches the R22 results much more closely than the blue posterior. Even when using a smaller subset of the observational data, the blue posteriors remain narrowly peaked in the parameter space, while the red posteriors become broader. The narrowness of the blue posteriors indicates that the blue broadband data determines the overall shape of the posteriors in the full broadband data inference.

\begin{figure}
    \centering
    \includegraphics[width=\columnwidth]{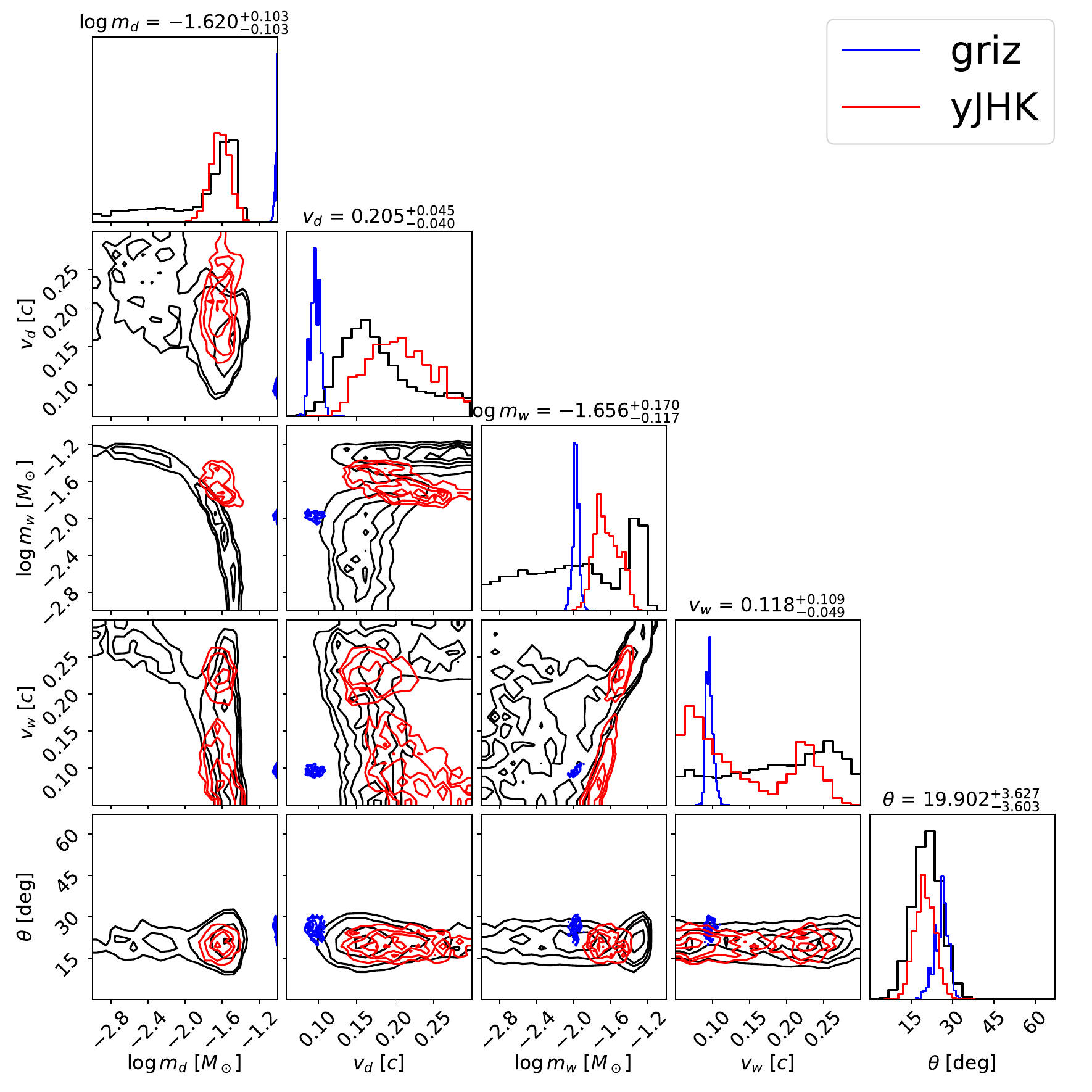}
    \caption{\edited{Two separate parameter inference runs, with the first (in blue) using data from only the \emph{griz} bands and the second (in red) using data from only the \emph{yJHK} bands. The red posteriors match the R22 posteriors (in black) much more closely. Parameter values at the top of each column represent the median \emph{yJHK} posterior values.}}
    \label{fig:blue_vs_red}
\end{figure}

The seemingly disproportionate effect of the blue broadband data on the posteriors could be attributed to the rapid evolution of the bluer bands compared to the red bands. As can be seen in Figure~\ref{fig:lc}, the evolution of the \emph{griz} light curves is more rapid than the \emph{yJHK} bands, with the \emph{griz} light curves dimming by $\sim 4$ magnitudes compared to the \emph{yJHK} bands dimming by two magnitudes during the first 10 days of observations. As such, the \emph{griz} light curves will be more restrictive regarding which model parameters fit the data, which is evident from the Figure~\ref{fig:blue_vs_red} posteriors.}

\subsection{Discussion}
\label{sec:discussion}
To summarize, following R22 we fit the same simulations and performed comparable inference of AT2017gfo.  After assessing
our training data and fit systematics, we adopted more conservative systematic uncertainties than R22.  We nonetheless
find dramatically narrower posteriors than R22, with inferred light curves consistent with observational
predictions. 

Several possible reasons for the qualitative discrepancies arise, primarily pertaining to our systematic uncertainty
estimate.  In R22, the Gaussian process interpolation  provided a pointwise and parameter-dependent error estimate,
which we qualitatively verified across the parameter space.  Also, the fitting strategy adopted in R22 independently fit each
timestep.   As a result, we expect that R22's models are unlikely to have correlated systematics in time, angle, and
wavelength.  For example, the R22 light curves occasionally have small but notable random discontinuities, consistent
with their reported fitting uncertainty.   By contrast, in this work, we fit all times and angles together within the
same training set \edited{to generate a vector prediction}. 
Our approach does not presently provide a pointwise error estimate.  Therefore, for the method
adopted in this work, we expect 
correlated errors in time and angle, but lack a method to characterize them versus those parameters or even the
intrinsic kilonova parameters.
\edited{Our MLP's vector light-curve prediction necessitates fitting the entire light curve for a certain band given a single set of parameters. In addition, as mentioned in Section~\ref{sec:blue_vs_red}, the blue bands are particularly constraining due to their more significant evolution over the observation period. Combining both of these effects results in only an extremely narrow region of the parameter space fitting all broadband data consistently. We did not encounter such narrow posteriors in R22 as each prediction was made for a specific time point; as such, many samples could reasonably fit an observation at any given time, resulting in broader posteriors when all times were stitched together to form the light curve.}

\edited{We use the validation loss value to roughly estimate the systematic fitting error associated with our neural network outcomes: \edited{the validation curve in Figure~\ref{fig:loss}} suggests that differences of order $\sqrt{MSE}\simeq \sqrt{0.2}$ magnitudes should occur in our predictions.  In practice, as illustrated in Figure~\ref{fig:sigma_sys} below, we find that the
average squared systematic error suggests a larger value than the validation MSE.}
We therefore anticipate that our naive estimate of $\sigma_{\rm sys}=0.5$, though well motivated by our detailed
followup study, \edited{may still} understate the systematic uncertainty inherent in our fitting approach, resulting in
narrow posterior distributions.
\edited{We also emphasize that the differences are not simply a matter of scale: the investigations performed in Figure~\ref{fig:sigma_sys} suggest that larger 
  white-noise systematic error cannot reconcile  differences between our current analysis and previous results.}

\edited{We note that we have experienced similar systematic uncertainty associated with observations in blue bands in R22 and an associated inference using simulations of spectra \cite{2023PhRvR...5d3106R}. The systematics in those works were related to our inability to reproduce the observed blue flux at times past $\sim 2$ days using our best-fit simulations. The systematics in this work, though similar in their connection to blue observations, introduce slightly different effects in our resultant inference. As we solve the bigger problem of matching our simulations to late-time blue observations, we anticipate that a more sophisticated treatment of our emulator systematics will allow us to better understand the effects of blue-band data on our inferences.}

A thorough investigation of suitable fit systematics for this \edited{neural network} is well beyond the scope of
our study. \edited{In the meantime, the neural network is suitable for investigations such as the one presented in Figure~\ref{fig:blue_vs_red}, where we can examine our models' ability to fit certain subsets of the data. In the \emph{griz} case, we see that our models require over 0.1 $M_\odot$ of slow-moving dynamical ejecta to fit the blue data. But, we  expect dynamical ejecta to be less massive and faster, thus potentially suggesting a missing modeling component}.

\section{Conclusions}
\label{sec:conclusion}


We present a \edited{neural network} architecture that is useful for the interpolation of kilonova light curves. We report on the \edited{network's} training and validation loss as a metric of successful training, as well as present examples of off-sample light-curve recovery. We use the \edited{neural network} to infer the parameters of the AT2017gfo kilonova and compare to previous inference performed in Ref.~\cite{Ristic22}. We find that the inference results are quite different from those previously obtained, but the light curves generated by the recovered parameters align well with the observational data. In particular, we investigate the \edited{neural network's} ability to seemingly infer narrow regions of the angle space despite being trained on light-curve data that should not allow for such specific inference. Given a detailed analysis of the mismatch between the \edited{neural network's} predictions and a simulation with higher-resolution angular data, we find that the \edited{network's pointwise systematic errors are consistent with our error estimate. However, our investigations also suggest that the systematic errors are correlated, not independent, in time and angle, in a way that is not captured by our model for systematic uncertainties.} In other words, we have discovered that the \edited{neural network's} goodness-of-fit varies appreciably across the time-angle space. While some of these variations are expected, others form interesting features that we cannot readily explain. We leave the analysis of the interpretability of these features for a future investigation.

\edited{We also show that the systematic uncertainty may be more complex than assumed in our simple uncorrelated (white
  noise) error model.  This was not the case in R22 due to the interpolation uncertainty, which naturally stemmed from the Gaussian process methodology. In recovering different parameters for AT2017gfo using two emulators trained on the same library of simulations, we highlight the importance of quantifiable uncertainty analysis in using emulators for robust inference.
As we do not present a way to handle correlated uncertainties in this work, a detailed uncertainty analysis, along with the resultant effects on parameter inference, will be necessary in future work.}

\section{Acknowledgments}
ROS and MR acknowledge support from NSF AST 1909534 and AST 2206321. AK also acknowledges support from NSF AST 2206321. VAV acknowledges support by the NSF through grant AST-2108676. The work by CLF, CJF, MRM, OK, and RTW was supported by the US Department of Energy through the Los Alamos National Laboratory (LANL). MRM acknowledges support from the Directed Asymmetric Network Graphs for Research (DANGR) initiative at LANL. This research used resources provided by LANL through the institutional computing program. Los Alamos National Laboratory is operated by Triad National Security, LLC, for the National Nuclear Security Administration of U.S.\ Department of Energy (Contract No.\ 89233218CNA000001).

\appendix

\section{Validating inference method}

\begin{figure}
    \centering
    \includegraphics[width=\linewidth]{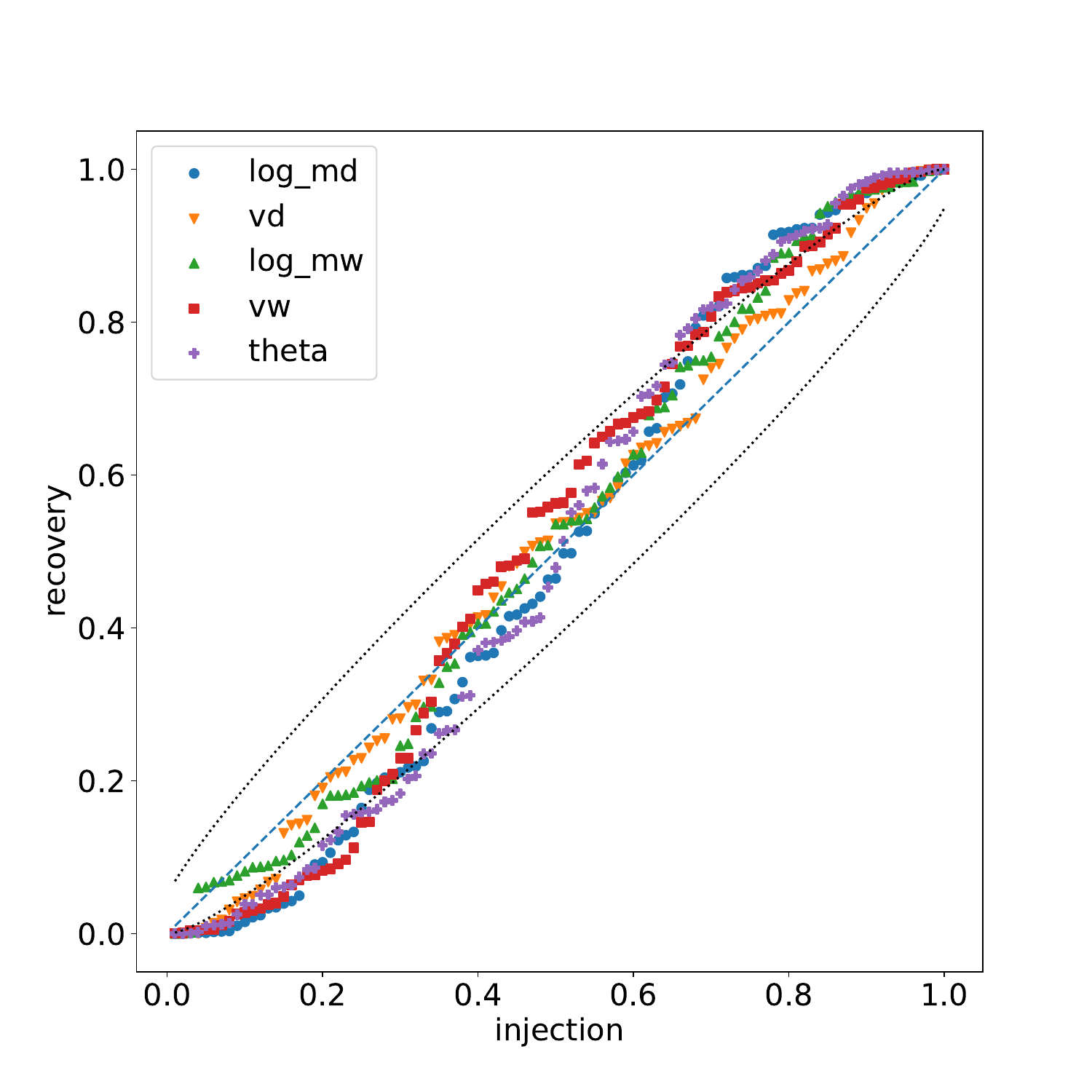}
    \caption{\edited{Probability-probability plot to validate recovery of
synthetic kilonova light curves, following  \cite{gwastro-RIFT-Update}.   Points show the PP plot data, with each simulation
parameter indicated as in the legend.  The dotted curves show the expected value and estimated 90\% credible binomial
interval.}}
    \label{fig:pp_plot}
\end{figure}

\edited{
To validate the statistical purity of our brute-force inference technique and our understanding of the noise model, we constructed a
standard  probability-probability (PP) plot test \cite{mm-stats-PP,gwastro-skyloc-Sidery2013}.
Our description follows the notation and narrative used in Ref. \cite{gwastro-RIFT-Update}.
For each source $k$, with true parameters $\mathbf{\lambda}_k$, we calculate
the fraction of  its posterior distribution with parameter $\lambda_\alpha$  below the true source value $\lambda_{k,\alpha}$   [$\hat{P}_{k,\alpha}(<\lambda_{k,\alpha})$].  After reindexing the sources so that $\hat{P}_{k,\alpha}(\lambda_{k,\alpha})$ increases with $k$ for some fixed $\alpha$,  a
plot of $k/N$ versus $\hat{P}_k(\lambda_{k,\alpha})$ can be compared with the expected diagonal result
($P(<p)=p$) and binomial uncertainty interval.
Figure~\ref{fig:pp_plot} shows the PP plot derived using kilonova light curves generated with our neural network
interpolator.  In these analyses, we adopt precisely the same observation cadence and uncertainties as AT2017gfo.  As in
our fiducial analysis of AT2017gfo, we adopt $\sigma_{\rm sys}=0.5$.  Each synthetic observation incorporates both
observational and (white noise) systematic uncertainty, added in quadrature consistent with our assumed likelihood.

The PP plot in Figure~\ref{fig:pp_plot}, being sufficiently consistent with the binomial credible interval, suggests that the brute-force Monte Carlo
inference strategy adopted in this work suffices for our purposes: in short, that the qualitative extent and character of
the posteriors shown in our figures are reasonably correct, such that the considerable tension between our analysis and
previous inferences accurately reflects the posterior.   We have specifically chosen to present results from a
brute-force inference technique to circumvent debates about our choice of implementation or our method of assessing
convergence.   With AT2017gfo, we have confirmed that the choice of integrator also doesn't qualitatively change our
answer:  alternative brute-force Monte Carlo integrator implementations
produce similar results.   That said, the PP
plot above is clearly not as diagonal as would expected for a well-developed and calibrated Bayesian inference algorithm
applied to this problem: its S-shape
features suggests either
modest overdispersion in our synthetic error model or modest underdispersion in our posterior distributions.
}

\bibliography{yinglei}

\end{document}